\documentclass[aps,pre,twocolumn,showpacs,showkeys]{revtex4-1}

\usepackage{graphicx}
\usepackage{amssymb}
\usepackage{amsmath}
\usepackage{bm}
\begin{document}
\title{Scaling of Nestedness in Complex Networks}
\author{Deok-Sun \surname{Lee}}
\email{deoksun.lee@inha.ac.kr}
\affiliation{Department of Natural Medical Sciences and Department of Physics, Inha University, Incheon 402-751}
\author{Seong Eun \surname{Maeng}}
\affiliation{Department of Physics, Inha University, Incheon 402-751}
\author{Jae Woo \surname{Lee}}
\affiliation{Department of Physics, Inha University, Incheon 402-751}


\begin{abstract}
Nestedness characterizes the linkage pattern of networked systems, indicating the likelihood that a node is linked to the neighbors of the nodes with larger degrees than it. Networks of mutualistic relationship between distinct groups of species in ecological communities exhibit such nestedness, which is known to support the network's robustness. Despite such importance, the quantitative characteristics of nestedness are little understood.  Here, we take a graph-theoretic approach to derive the scaling properties of nestedness in various model networks.  Our results show how the heterogeneous connectivity patterns enhance nestedness. Also, we find that the nestedness of bipartite networks depends sensitively on the fraction of different types of nodes, causing nestedness to scale differently for nodes of different types.
\end{abstract}


\maketitle 

\section{Introduction}
The patterns of interspecific interaction determines the structure and the evolution of ecological networks~\cite{pascual05,williams00,cattin04,allesina08}. Among them, the mutualistic relationship between distinct groups of species such as flowering plants and pollinating animals is represented by bipartite networks, the topological feature and stability of which have received much attention recently~\cite{bascompte03,bascompte06, montoya06,saavedra09,bastolla09,bascompte10,hwang08,maeng11a}.  Nestedness is a remarkable feature of mutualistic networks, which means that specialists - the species interacting with a small number of other species - tend to interact with the species that interact with generalists - the species interacting with a large number of species~\cite{bascompte03}. Such nestedness is manifested in the adjacency matrix of a given network. If rows and columns are arranged from the most generalist to the most specialist, the adjacency matrix has $1$'s filling the upper-left corner for networks with perfect nestedness. A couple of measures have been introduced to quantify nestedness, including the matrix temperature representing the deviation of the adjacency matrix from the perfect nestedness benchmark~\cite{atmar93} and the mean topological overlap between nodes~\cite{neto08}, the latter of which is used in this work.

There have been many studies on the model for mutualistic networks~\cite{jordano03,vazquez05,guimaraes07,medan07,maeng11b}. Furthermore, nestedness was shown to underlie the robustness of mutualistic networks, in contrast to trophic networks displaying modularity for the stability of the intertwined predator-prey relationship~\cite{thebault10}. Despite its universality and such profound impact on the function and the evolution of ecological systems, the quantitative characteristics of nestedness, such as the scaling behavior of nestedness with system size and its dependence on network structure, are little understood. In this paper, we take a graph-theoretic approach to study the scaling property of nestedness in networks of the static model~\cite{goh01,lee04,jslee06}, the Barab\'{a}si-Albert (BA) model~\cite{barabasi99}, and the BA-type bipartite network model~\cite{maeng11b}. We investigate the mean topological overlap between nodes as the measure of nestedness~\cite{neto08}. We find that nestedness vanishes in the thermodynamic limit. However, for finite systems, the heterogeneity of node connectivity, a universal feature of real-world networks, can enhance nestedness such that its scaling properties are changed in the case of strongly heterogeneous networks. Also, we show how the nestedness in bipartite networks depends on the ratio of the numbers of different types of nodes and their evolution rules. 

The paper is organized as follows: In Sec.~\ref{sec:measure}, we introduce the concept and the measure of nestedness. Then, we derive the nestedness of unipartite networks in the static model and the BA model in Sec.~\ref{sec:uni}. The BA-type bipartite network model is introduced and its nestedness is studied in Sec.~\ref{sec:bi}. Our findings are summarized and discussed in Sec.~\ref{sec:discussion}. In Appendix~\ref{sec:appendix_static}, we discuss in more detail the computation of nestedness in the static model. 

\section{Measure of nestedness}
\label{sec:measure}

\begin{figure}
\includegraphics[width=0.9\columnwidth]{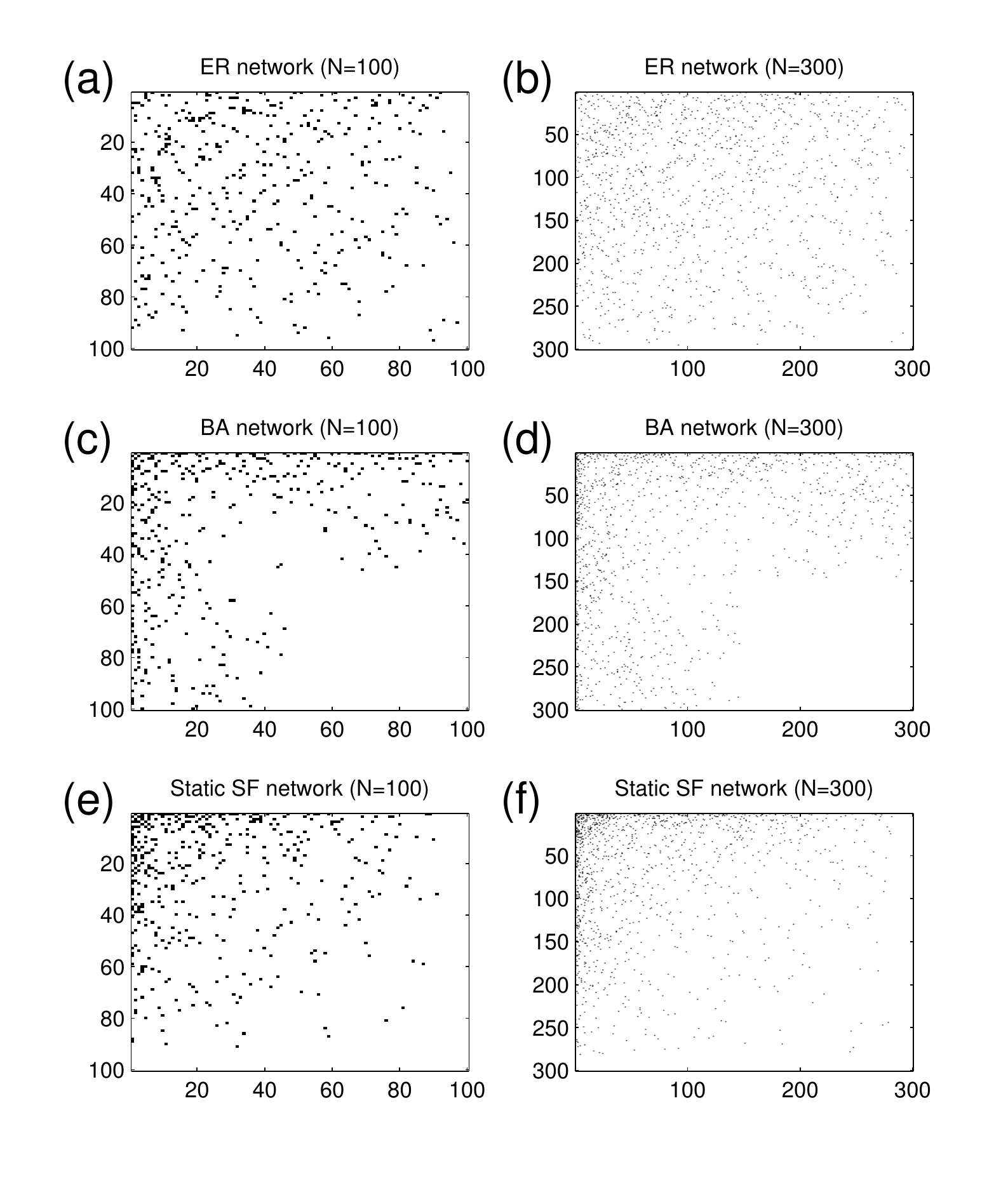}
\caption{Adjacency matrices of randomly-linked and scale-free networks. Node labels are assigned in descending order of degree. All the networks have the same value of  ${L\over N} = 2$, with $N$ being the number of nodes and $L$ being the number of links. (a) ER network with $N=100$ and $L=200$. (b) ER network with $N=300$ and $L=600$. (c) BA network with $N=100$ and $m=2$. (d) BA network with $N=300$ and $m=2$. (e) Scale-free network in the static model with $N=100, L=200$ and $\gamma=2.4$. (f) Scale-free network in the static model with $N=300, L=600$ and $\gamma=2.4$.}
\label{fig:adj}
\end{figure}

The nestedness of a network represents the likelihood of a node being linked to the neighbors of nodes that have larger degrees than it. In perfectly nested networks, if the degree - number of neighbors - of node $i$ is smaller than that of node $\ell$, the neighbors $\mathcal{N}_i$ of node $i$ are necessarily a subset of $\mathcal{N}_\ell$, the set of the neighbors of node $\ell$~\cite{bascompte03}. 

Adjacency matrices of nested networks show some peculiarities. Suppose that the node indices $i=1,2,\ldots,N$  are assigned such that the degrees of two nodes $i$ and $j$ satisfy $k_i \leq k_j$ for $i>j$. Then, $1$'s are expected to appear more in the upper-left corner of the adjacency matrix $a_{ij}$. For networks with perfect nestedness, $a_{i\ell}=1$ implies that $a_{j\ell}=1$ for all $j\leq i$; therefore, the adjacency matrix has $1$'s filling  the upper-left corner compactly. The matrix temperature, an original measure of nestedness, measures the difference in the distribution of $1$'s and $0$'s in the adjacency matrix from that in a network with perfect nestedness~\cite{atmar93}. In the adjacency matrices shown in Fig.~\ref{fig:adj},  dots representing $1$'s are found more in the upper-left corners to a variable extent.  Figures~\ref{fig:adj}(a) and (b) represent the adjacency matrices of randomly-linked networks or Erd\"{o}s-R\'{e}nyi  (ER) networks~\cite{albert02} and Figs.~\ref{fig:adj} (c)-(f) are those of the scale-free (SF) networks with power-law degree distributions $p_d(k)\sim k^{-\gamma}$~\cite{barabasi99,goh01,lee04,jslee06,albert02}. The tendency to fill dots compactly in the upper-left corners is shown to be more significant in scale-free networks and for smaller system size. Characterizing quantitatively such impacts of network structure and system size on nestedness is the purpose of the presented study. 

The calculation of the matrix temperature of a network is not simple, making it hard to understand the behavior of nestedness~\cite{atmar93,bascompte03,neto08}. Recently, a simple and intuitive measure was introduced, the performance of which was shown to be as good as or better than the matrix temperature~\cite{neto08}. Following Ref.~\cite{neto08}, we define the nestedness of a unipartite network of $N$ nodes with the adjacency matrix $a_{ij}$ as  
\begin{equation}
S = {1\over N(N-1)} \sum_{i=1}^N \sum_{j=1}^N  {\sum_{\ell=1}^N a_{i\ell} a_{j\ell}\over \min (k_i,k_j)},
\label{eq:S}
\end{equation}
where $k_i =\sum_{\ell=1}^N a_{i\ell}$ is the degree of node $i$. For nodes $i$ and $j$, the ratio of the numbers of their common neighbors, $\sum_\ell a_{i\ell}a_{j\ell}$, to their minimum degree is identical to their topological overlap, which has been used widely to quantify the similarity of two nodes in biological and social networks~\cite{ravasz02}. Nestedness $S$ is then equal to the mean topological overlap averaged over all pairs of nodes. $S$ in Eq.~(\ref{eq:S}) is consistent with the original concept of nestedness, the tendency that the neighbors of a node with smaller degree belong to the set of neighbors of another node with larger degree. It is straightforward to extend Eq.~(\ref{eq:S})  to the case of bipartite networks, which will be given in Sec.~\ref{sec:bi}. 

For an ensemble of networks of $N$ nodes indexed by $i=1,2,\ldots,N$, the probability for two nodes $i$ and $j$ to be connected is given by $f_{ij}=\langle a_{ij}\rangle$, with $a_{ij}$ being the adjacency matrix elements and $\langle \cdot \rangle$ the ensemble average.  Suppose that the node indices  are assigned so that $\langle k_i\rangle \leq \langle k_j\rangle$ for $i>j$. Then, the ensemble average of  nestedness  is 
\begin{eqnarray}
S &=& \left\langle {1\over N(N-1)} \sum_{i=1}^N \sum_{j=1}^N  {\sum_{\ell=1}^N a_{i\ell} a_{j\ell}\over \min (k_i,k_j)}\right\rangle\nonumber\\
&=& {2  \over N(N-1)} \sum_{i=1}^N \sum_{j=1}^{i-1}  {\sum_{\ell=1}^N  f_{i\ell} f_{j\ell}\over \langle k_i \rangle}.
\label{eq:S2}
\end{eqnarray}
Here, we assumed that the fluctuation of degree is much less than that of the adjacency matrix elements  and that  distinct adjacency matrix elements $a_{i\ell}$ and $a_{j\ell}$ were statistically independent. Under these assumptions, stochastic variables such as $a_{ij}$ and $k_i$ are replaced by their ensemble-averaged values in Eq.~(\ref{eq:S2}). 

If the connection probability $f_{i\ell}$  is of the form~\cite{park03}
\begin{equation}
f_{i\ell} = 2L P_i P_\ell,
\label{eq:factorized}
\end{equation}
where $L$ is the ensemble average of the total number of links and the node-selection probability $P_i$'s satisfy $0\leq P_i\leq 1$ and $\sum_{i=1}^N P_i =1$, then the expected degree $\langle k_i\rangle$ is simply given by 
\begin{equation}
\langle k_i \rangle =\sum_{\ell=1}^N f_{i\ell}=  2L P_i,
\end{equation}
and the nestedness $S$ is obtained from Eq.~(\ref{eq:S2}) as  
\begin{equation}
S = {2L\over N(N-1)}\sum_{i=1}^N \sum_{j=1}^{i-1} \sum_{\ell=1}^N P_j P_\ell^2  =  
{2L I_1 I_2 \over N(N-1)},
\label{eq:S3}
\end{equation}
with 
\begin{equation}
I_1 = \sum_{i=1}^N \sum_{j=1}^{i-1}P_j \ {\rm and} \ \   
I_2 = \sum_{\ell=1}^N  P_\ell^2.
\label{eq:I}
\end{equation}
Because we index nodes in decreasing order of the expected degrees, $P_\ell$ is a non-increasing function of $\ell$. We remark that the scaling behavior of $I_2$ depends crucially on how fast $P_\ell$ decays with $\ell$, which in turn determines the scaling of $S$. If all nodes have the same selection probability  $P_i =N^{-1}$, true for ER networks,  then $I_1 = {N\over 2}$, $I_2 ={1 \over N}$, and 
$S = {2L\over N^2}$. We consider in this work the case of ${L\over N}$ being finite, $\mathcal{O}(1)$; thus, $S\sim N^{-1}$~\cite{LS}.

The factorized form of the connection probability in Eq.~(\ref{eq:factorized}) is not always valid, but as we will see, can be used for the networks studied in this work, as far as the scaling behavior of nestedness  is concerned.  We focus on the behavior of $S$ given by 
\begin{equation}
S\simeq s_o  N^\theta \ \ {\rm for} \ N\to\infty,
\label{eq:Sasym}
\end{equation}
where $s_o$ is a constant and $\theta$ is the scaling exponent. In the next sections, we derive the scaling exponent $\theta$ analytically and check it against numerical simulations.

\section{Nestedness of unipartite networks}
\label{sec:uni}
\subsection{Static Model}
\label{sec:static}

In this section, we compute the nestedness of the static model for equilibrium scale-free networks~\cite{goh01,lee04,jslee06}. The model is a generalization of ER networks, incorporating the heterogeneous connectivity patterns identified in numerous complex systems. In the model, there are $N$ nodes indexed by $i=1,2,\ldots, N$ with no link initially. At each time step, two nodes $i$ and $j$ are selected with probability $P_i$ and $P_j$ with 
\begin{equation}
P_i = {i^{-\alpha}\over\zeta_N(\alpha)}
\label{eq:Pi}
\end{equation}
and connected if they are disconnected. Here, $0\leq \alpha<1$ and $\zeta_N(\alpha)=\sum_{i=1}^N i^{-\alpha}$. When a network evolves to time step $NK$, with $K$ being a constant, the expected number of links $L$ is $NK$, and the degree distribution has a power-law form; asymptotically $p_d(k)\sim k^{-\gamma}$ with the degree exponent $\gamma=1+{1\over\alpha}$~\cite{goh01,lee04,jslee06}. With $\alpha=0$, ER networks are generated. 

The connection probability $f_{i\ell}$ for $i\ne \ell$ in the static model is given by~\cite{lee04,jslee06}
\begin{equation}
f_{i\ell}=
1 - \left(1 - 2 P_iP_\ell\right)^{NK}=
1-e^{-2NK P_i P_\ell},
\label{eq:f}
\end{equation} 
where we used the relation $(1-x)^N = 1-e^{-Nx}$ for $N\to\infty$ and $x<1$. 
While $2NKP_i P_\ell$ is vanishingly small for $0\leq \alpha <{1\over 2}$ ($\gamma=1+{1\over\alpha}>3$), it can diverge, depending on $i$ and $\ell$, if ${1\over 2}\leq \alpha<1$, leading to $f_{i\ell}\simeq 1$, which  underlies the negative degree-degree correlations in scale-free networks with $2<\gamma\leq 3$~\cite{park03,jslee06}. 

When  $2NKP_i P_\ell\ll 1$, the connection probability $f_{i\ell}$  is factorized  as 
\begin{equation}
f_{i\ell} = 2NK P_i P_\ell.
\label{eq:factorized_static}
\end{equation}
If $0\leq \alpha<{1\over 2}$,  Eq.~(\ref{eq:factorized_static}) holds because $\max_{i,\ell}f_{i\ell}={2NK\over \zeta_N(\alpha)^2} = \mathcal{O}(N^{2\alpha-1})$. Then, we can insert $I_1$ and $I_2$ of Eq.~(\ref{eq:I}),
\begin{eqnarray}
I_1 &\simeq & \sum_{i=1}^N {i^{1-\alpha} \over (1-\alpha)\zeta_N(\alpha)} = {N\over 2-\alpha}, \nonumber\\
I_2 &=& \sum_{\ell=1}^N {\ell^{-2\alpha} \over \zeta_N(\alpha)^2} = {(1-\alpha)^2 \over (1-2\alpha)N},
\label{eq:I_static}
\end{eqnarray}
where we used $\zeta_N(\alpha)\simeq {N^{1-\alpha}\over 1-\alpha}$ for $0\leq \alpha<1$ and $N\gg 1$, into Eq.~(\ref{eq:S3}) to obtain 
\begin{equation}
S = {4K (1-\alpha)^2 \over (1-2\alpha)(2-\alpha)} N^{-1} \ {\rm for} \ 0\leq \alpha<{1\over 2}.
\label{eq:S_static_case1}
\end{equation}
This shows that in the thermodynamic limit $N\to\infty$ and ${L\over N}=\mathcal{O}(1)$, the nestedness $S$ vanishes. For finite systems, however, $S$ is non-zero, and its dependence on the network structure can be observed. As will be shown below, the nestedness decays slower than $N^{-1}$ in the case of ${1\over 2}\leq \alpha<1 (2<\gamma \leq 3)$. 

If ${1\over 2}\leq \alpha<1$, Eq.~(\ref{eq:factorized_static}) holds only for $\ell\gg c(i) = \left({2NK\over \zeta_N(\alpha)^2}\right)^{1\over \alpha} {1\over i}=\mathcal{O}\left(N^{2-{1\over \alpha}} {1\over i}\right)$. Nevertheless, the dominant contribution to the leading behaviors of the expected degree and the nestedness  in Eq.~(\ref{eq:S2}) is made by the terms with $f_{i\ell}$ approximated by the form in Eq.~(\ref{eq:factorized_static}). The details are given in Appendix~\ref{sec:appendix_static}. In this case, the value of $I_1$  is the same as Eq.~(\ref{eq:I_static}), and $I_2$ is evaluated as
\begin{equation}
I_2 =  \sum_{\ell=1}^N {\ell^{-2\alpha} \over \zeta_N(\alpha)^2} = \left\{
\begin{array}{ll}
\zeta(2\alpha) (1-\alpha)^2 N^{2\alpha-2} & \ {\rm for} \ \alpha>{1\over 2}\\
{\ln N\over 4N} & \ {\rm for} \ \alpha={1\over 2},
\end{array}
\right.  
\end{equation}
where $\zeta(2\alpha)$ is the Riemann-zeta function. Nestedness is then given by  
\begin{equation}
S = \left\{
\begin{array}{ll}
{4K (1-\alpha)^2 \over 2-\alpha}\zeta(2\alpha) N^{-(2-2\alpha)} & \ {\rm for} \ \alpha>{1\over 2}\\
{2K \over 3}{\ln N \over  N} & \ {\rm for} \ \alpha={1\over 2}.
\end{array}
\right.
\label{eq:S_static_case2}
\end{equation}

From Eqs.~(\ref{eq:S_static_case1}) and (\ref{eq:S_static_case2}), one sees the scaling exponent $\theta$ given by  
\begin{equation}
\theta = \left\{
\begin{array}{ll}
-1 & \ {\rm for} \ 0\leq \alpha <{1\over 2}  (\gamma>3)\\
-2(1-\alpha)=-2{\gamma-2\over \gamma-1}  & \ {\rm for} \ {1\over 2}<\alpha<1 (2<\gamma<3).
\end{array}
\right.
\label{eq:theta}
\end{equation}
Our results show first that the nestedness decreases with increasing system size in the static model without regard to their degree exponents. The decrease of nestedness with increasing system size is also seen in the adjacency matrices in Fig.~\ref{fig:adj}. Secondly, the nestedness is much higher in strongly heterogeneous networks, those with $2<\gamma\leq 3$, for finite system size. The hub nodes, which are rich in scale-free networks, act as 'common' neighbors of many pairs of nodes and, thus, enhance their topological overlap, giving rise to the slow decay of nestedness with system size for $2<\gamma\leq 3$.  One can see a more uneven distribution of non-zero elements in the adjacency matrix of scale-free networks with $\gamma=2.4$ than in the ER networks in Fig.~\ref{fig:adj}.

\begin{figure}
\includegraphics[width=0.9\columnwidth]{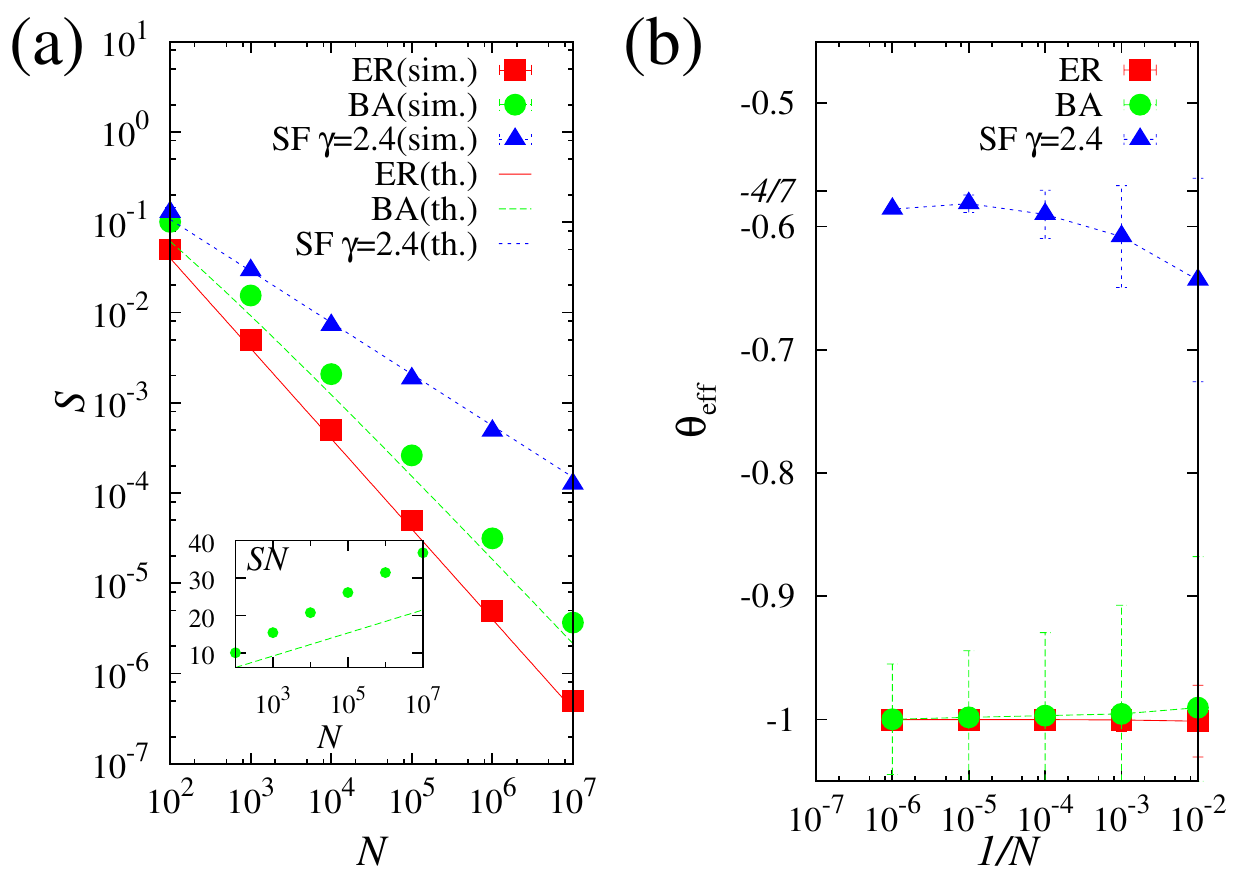}
\caption{Scaling of nestedness in unipartite networks. (a) Plot of $S$ versus system size $N$ for ER networks ($\alpha=0, K=2$) and  SF networks ($\alpha={5\over 7}, K=2$) of the static model and for BA networks ($m=2$) on a logarithmic scale. Data points are from simulation results, and lines are from analytic results. The inset shows plots of $SN$ versus $N$ for BA networks on a semi-log scale. While the behavior $SN\sim \ln N$ is identified, its coefficient deviates from the theoretical prediction (line). (b) Effective exponent $\theta_{\rm eff}$ obtained numerically from simulation results plotted as a function of the inverse of the system size $1/N$. The analytic results $\theta=2\alpha-2 = -{4\over 7}$ for SF networks ($\alpha={5\over 7}$) and $\theta=1$ for ER ($\alpha=0$) and BA networks are also shown for comparison. The error bars represent the standard deviation.}
\label{fig:uni}
\end{figure}

To check these analytic predictions, we performed simulations of the static model for different system sizes $N$ and $\alpha$ (see Fig.~\ref{fig:uni}).  The decay of nestedness with increasing system size is indeed much slower for $\alpha={5\over 7} (\gamma=2.4)$ than for $\alpha=0$ ($\gamma\to\infty$; ER network). With the numerical results, we determine the effective exponent $\theta_{\rm eff}$ as
\begin{equation}
\theta_{\rm eff}(N) = {\ln \left({S(N')\over S(N)}\right) \over \ln \left({N'\over N}\right)},
\label{eq:thetaeff_static}
\end{equation}
where $S(N)$ is the nestedness obtained for networks of size $N$ generated in the simulations and $N'=10N$ was used. Figure.~\ref{fig:uni}(b) shows that $\theta_{\rm eff}$ converges to the theoretical values as $N$ increases. On the other hand,  the coefficient $s_o$ in Eq.~(\ref{eq:Sasym}) seems to show a deviation between the theory and the numerics, probably due to the approximation of the connection probability by the factorized form in Eq.~(\ref{eq:factorized_static}).  Actually, the coefficients in  Eqs.~(\ref{eq:S_static_case1}) and (\ref{eq:S_static_case2}) correspond to the upper bound of $s_o$, as shown in Appendix~\ref{sec:appendix_static}.

\subsection{BA Model}
\label{sec:ba}
 
In this section, we consider the BA model for growing scale-free networks~\cite{barabasi99}. In the model,  initially there are $m+1$ nodes that are fully connected. At each time step, a new node arrives, and $m$ old nodes are selected with probability proportional to their degrees and connected to the new node. Repeating this procedure  up to the $N$th time step, we obtain a realization of the BA network of $N$ nodes and $Nm$ links, neglecting the $m$ initial nodes and their $m(m-1)/2$ links for $N$ large. The set of many such realizations forms an ensemble. If we index the nodes by their arrival time steps $i=1,2,\ldots,N$, node $i$ is connected to node $\ell(>i)$ with probability $m{k_i(\ell)\over \sum_{j=1}^\ell k_j(\ell)}$; thus, the connection probability $f_{i\ell}$ in the network ensemble is given by 
\begin{equation}
f_{i\ell} =\left\langle m { k_i(\ell)\over\sum_{j=1}^{\ell} k_j(\ell)}\right \rangle = {\langle k_i(\ell)\rangle \over 2 \ell}, 
\end{equation}
where $\langle k_i(\ell)\rangle$ is the expected degree of node $i$ at time step $\ell$.  $\langle k_i(t)\rangle$ evolves with time as ${d\over dt}\langle k_i(t)\rangle  =f_{it} = {\langle k_i(t)\rangle\over 2t}$ for $t\geq i$, and we obtain 
\begin{equation}
\langle k_i(\ell)\rangle= m\sqrt{\ell\over i}.
\label{eq:kBA}
\end{equation}
The degree distribution, $p_d(k)\sim k^{-3}$, can be derived from Eq.~(\ref{eq:kBA})~\cite{albert02}. The connection probability can be  brought into the factorized form 
\begin{equation}
f_{i\ell} = {m\over 2\sqrt{i\ell}} = 2L P_i P_\ell
\label{eq:fBA}
\end{equation} 
with the node-selection probability 
\begin{equation}
P_i = {i^{-{1\over 2}} \over \zeta_N({1\over 2})}
\label{eq:PiBA}
\end{equation}
satisfying $\sum_{i=1}^N P_i = 1$
and $L = {m\over 4 \zeta_N(1/2)^2}\simeq Nm$. Note that $f_{i\ell}$ is the connection probability in the ensemble of BA networks grown up to time step $N$. 

The node-selection probability and the connection probability in Eqs.~(\ref{eq:fBA}) and (\ref{eq:PiBA}) are identical to those of the static model with $\alpha={1\over 2}$ given in Eqs.~(\ref{eq:Pi}) and (\ref{eq:f}) except for the replacement of $K$ by $m$. Therefore, the nestedness of the BA model shows the same scaling as that of the static model with $\alpha={1\over 2}$, that is, 
\begin{equation}
S = {2m \over 3} {\ln N\over N}.
\label{eq:S_BA}
\end{equation} 
By the computer simulations of the BA model, we determined numerically the nestedness of BA networks, which is consistent with Eq.~(\ref{eq:S_BA}) as shown in Fig.~\ref{fig:uni}. Under the assumption $S\sim \left({N\over \ln N}\right)^{\theta_{\rm eff}}$, the effective exponent $\theta_{\rm eff}$ is numerically determined by
\begin{equation}
\theta_{\rm eff}(N) = {\ln \left({S(N')\over S(N)}\right) \over \ln \left({N'\ln N\over N\ln N'}\right)},
\label{eq:thetaeff_BA}
\end{equation}
with $N'=10N$,  which converges to $1$ as in Fig.~\ref{fig:uni} (b).  The coefficient $s_o$ in the relation $S\simeq s_o {\ln N\over N}$ seems to  deviate from the analytic prediction $2m/3$ (see Fig.~\ref{fig:uni}(a)). The deviation seems to originate in the dynamical  correlation between  $a_{i\ell}$ and $a_{j\ell}$, which was neglected in deriving Eq.~(\ref{eq:S2}). While  $a_{i\ell}$ and $a_{j\ell}$ are independent in the static model, they can be positively correlated  in the BA model; for instance, if $ i<\ell<j $, the expected degree, $\langle k_\ell(j)\rangle$, of node $\ell$ at time $j$ is larger, and in turn the  probability for a new node $j$ to select node $\ell$ as its partner is larger, when nodes $i$ and $\ell$ are connected than when they are disconnected. 

\section{Nestedness of BA-type bipartite networks}
\label{sec:bi}

\subsection{Model}

Mutualistic networks such as plant-pollinator networks are bipartite networks consisting of two types of nodes, for example, animal ($A$) and plant ($P$) types. Each node represents a distinct species, and a link between two nodes of type  A and P indicates that the corresponding animal species pollinate the corresponding plant species, a beneficial interaction for their survival and reproduction. Given the broad degree distributions in real-world mutualistic networks~\cite{jordano03,guimaraes07,maeng11a} and diverse interspecific relationships among species~\cite{maeng11b}, one can consider an extension of the BA model for a bipartite structure to understand the evolution of bipartite heterogeneous networks~\cite{maeng11b,ramasco04,goldstein05,peruani07}. We consider the following BA-type bipartite network model. 
Initially,  there are $m_P$ nodes of type $P$ and $m_A$ nodes of type $A$. All pairs of nodes of unequal types are connected. At each time step, a node of type $P$ is newly introduced with probability $\rho_P$ or a node of type $A$ is introduced with probability $\rho_A=1-\rho_P$. The new node of type $P\  (A)$ selects   $m_P\ (m_A)$ nodes of type $A \ (P)$  with a probability proportional to their degrees and connects itself to them. Iterating these procedures up to time step $N$,  one obtains a bipartite network of $N\rho_P$ nodes of type $P$ and $N\rho_A$ nodes of type $A$, on the average. 

As in the unipartite case, we index each node by its arrival time $i=1,2,\ldots, N$.  If $i<\ell$ and node $i$ is of type $A$ and node $\ell$ is of type $P$, their connection probability  is given by 
\begin{equation}
 f_{i\ell}^{(AP)} =\left\langle m_P { k_i^{(A)}(\ell)\over \sum_{j=1}^{\ell} \delta(b_j,A) k_j^{(A)}(\ell)} \right\rangle
 = m_P {\langle k_i^{(A)}(\ell) \rangle\over \ell\langle m\rangle},
\label{eq:fbi0}
\end{equation}
where $\delta(a,b)$ is the Kronecker delta function,  $b_j = A$ or $P$ indicates the type of node $j$, and $\langle m\rangle = \rho_A m_A + \rho_P m_P$ is the mean number of initial links of new nodes. Also,  $\langle k_i^{(A)}(\ell)\rangle$ is the expected degree of type-$A$ node $i$ at time step $\ell$. If node $i$ is of type $P$ and node $\ell$ is of type $A$, their connection probability $f_{i\ell}^{(PA)}$ will be identical to Eq.~(\ref{eq:fbi0}), but with $A$ and $P$ exchanged. $\langle k_i^{(A)}(t)\rangle$ increases at time $t$ if the newly-arrived node is of type $P$ and connected to node $i$, which occurs with probability $\rho_P f_{it}^{(AP)}$. Therefore, the time-evolution equation of the expected degree $\langle k_i^{(A)}(t)\rangle$ is given by 
 \begin{equation}
{d \over dt}\langle k_i^{(A)}(t)\rangle  =  \rho_P f_{it}^{(AP)}
= {1\over t \mu_A } \langle k_i^{(A)}(t)\rangle,
\label{eq:kitBi}
\end{equation}
where we introduced the parameter $\mu_A = {\langle m\rangle\over \rho_P m_P}= 1+ {\rho_A m_A \over \rho_P m_P}$. Similarly, $\mu_P =  {\langle m\rangle\over \rho_A m_A}$. Solving Eq.~(\ref{eq:kitBi}), we obtain
\begin{equation}
\langle k_i^{(A)}(\ell)\rangle  = m_A \left({\ell\over i}\right)^{1\over\mu_A}.
\label{eq:kiBi}
\end{equation}
Similarly, the evolution of the expected degree of a type-$P$ node $i$ will be described by $\langle k_i^{(P)}(\ell)\rangle  = m_P \left({\ell\over i}\right)^{1\over\mu_P}$; thus, the values of $\mu_A$ and $\mu_P$ essentially determine the time-evolutions of $\langle k_i^{(A)}\rangle$ and $\langle k_i^{(P)}\rangle$. Using these results, one can derive the degree distributions $p_d^{(P)}\sim k^{-\gamma_P}$ and $p_d^{(A)}\sim k^{-\gamma_A}$ with $\gamma_P = 1+\mu_P$ and $\gamma_A = 1+\mu_A$~\cite{maeng11b}. Also, the connection probability $f_{i\ell}^{(AP)}$ is then represented as
\begin{equation}
f_{i\ell}^{(AP)} = {m_A m_P \over \langle m\rangle} {1\over  i^{1\over\mu_A} \ell^{1\over\mu_P}},
\label{eq:fBi}
\end{equation}
where we used the relation ${1\over\mu_A} + {1\over\mu_P} =1 $. From Eq.~(\ref{eq:fBi}), one finds that the connection probability between nodes $i$ and $j$ is not affected by whether $i<j$ or $i>j$, but depends only on their node types through $\mu_A$ and $\mu_P$.  $f_{i\ell}^{(AP)}$ in Eq.~(\ref{eq:fBi}) is given in the factorized form as 
\begin{equation}
f_{i\ell}^{(AP)} = L {P_i^{(A)}P_\ell^{(P)} \over \rho_A \rho_P},
\label{eq:factorized_BI}
\end{equation}
where $L= N\langle m\rangle$ is the expected number of links and 
\begin{equation}
P_i^{(A)} = {i^{-{1\over \mu_A}}\over \zeta_N({1\over \mu_A})} \ {\rm and} \ P_\ell^{(P)} = {\ell^{-{1\over \mu_P}}\over \zeta_N({1\over \mu_P})}.
\end{equation}
Then, the expected degree of type-$A$ nodes is 
\begin{equation}
\langle k_i^{(A)}(\ell)\rangle =  \sum_{j=1}^\ell \rho_P f_{i\ell}^{(AP)} = {L P_i^{(A)}\over\rho_A},
\label{eq:kABI}
\end{equation}
that of type-$P$ nodes is the same as the above with $A$ replaced by $P$,  
and it holds that $L = \sum_{i=1}^N \rho_A \sum_{\ell=1}^N \rho_P f_{i\ell}^{(AP)}=N\langle m\rangle$.  

\subsection{Nestedness}

In BA-type bipartite networks, the topological overlap between two $A$-type nodes is expected to depend on how many $P$-type nodes are present in the networks, which is controlled by $\rho_A$ and $\rho_P$.  In this section, we derive the nestedness of BA-type bipartite networks and investigate its dependence on the model parameters. 

For bipartite networks, the nestedness can be considered separately for  each  node-type. Let $b_i=A$ or $P$ be the node-type variable and $a^{(AP)}$ denote the $N\times N$ adjacency matrix with $a_{i\ell}^{(AP)} =1$ only when $b_i=A$, $b_\ell=P$, and $i$ and $\ell$ are connected. Then,  the nestedness $S^{(A)}$ for type-$A$ nodes can be defined as  
\begin{eqnarray}
S^{(A)} &=&{2\over  N_A(N_A-1)} \sum_{i=1}^{N} \delta(b_i,A) \sum_{j=1}^{N} \delta(b_j,A) \nonumber\\
&&  \sum_{\ell=1}^{N}\delta(b_\ell,P) 
{a^{(AP)}_{i\ell} a^{(AP)}_{j\ell}\over \min (k^{(A)}_i,k^{(A)}_j)}.
\label{eq:SBi}
\end{eqnarray}
Below, we will consider only $S^{(A)}$. All the results obtained for $S^{(A)}$ can be applied to $S^{(P)}$ after exchanging $A$ and $P$. Consider the ensemble of BA-type bipartite networks of $N$ nodes with $\rho_A, \rho_P, m_A$, and $ m_P$ given. The node indices represent their arrival times; therefore, $\langle k^{(A)}_i\rangle \leq \langle k_j^{(A)}\rangle$ and $\langle k^{(P)}_i\rangle \leq \langle k_j^{(P)}\rangle$ for $i>j$. Taking the ensemble average of the stochastic variables such as $\delta(b_\ell,A)$ and $a_{j\ell}^{(AP)}$ in Eq.~(\ref{eq:SBi}), we find  
\begin{equation}
 S^{(A)} ={2\over N\rho_A (N\rho_A-1)} \sum_{i=1}^N \rho_A \sum_{j=1}^{i-1} \rho_A \sum_{\ell=1}^N \rho_P{f_{i\ell}^{(AP)} f_{j\ell}^{(AP)}\over\langle k_i^{(A)}(N)\rangle}.
\label{eq:S2Bi}
\end{equation}
This is the bipartite version of Eq.~(\ref{eq:S2}). Using $f^{(AP)}$  given in Eq.~(\ref{eq:factorized_BI}), we can compute $S^{(A)}$ as 
\begin{equation}
S^{(A)} = {2 L \over N\rho_A (N\rho_A -1)} {\rho_A \over \rho_P} I_1^{(A)} I_2^{(P)},
\label{eq:S3_BI}
\end{equation}
with 
\begin{equation}
I_1^{(A)} = \sum_{i=1}^N\sum_{j=1}^{i-1} P_i^{(A)} \ {\rm and} \ 
I_2^{(P)} = \sum_{\ell=1}^N (P_\ell^{(P)})^2. 
\end{equation}

As already shown in Sec.~\ref{sec:static}, $I_1^{(A)}$ scales as $\sim N$ as given by Eq.~(\ref{eq:I_static}) with ${1\over\mu_A}$  in place of $\alpha$.  That is,
\begin{equation}
I_1^{(A)} \simeq {N \over 2-{1\over \mu_A}}. 
\label{eq:I1_BI}
\end{equation}
On the other hand, the scaling of $I_2^{(P)}$'s depends on whether $\mu_P$ is larger than $2$ or not. When  $\mu_P>2$,
\begin{equation}
I_2^{(P)} =  {1\over \mu_A^2 (1 - 2 {1\over \mu_P})} N^{-1},
\label{eq:I2_BI_case1}
\end{equation}
and when  $1\leq\mu_P<2 $,
\begin{equation}
I_2^{(P)} = {\zeta({2\over \mu_P})\over \mu_A^2 }N^{-{2\over \mu_A}}.
\label{eq:I2_BI_case2}
\end{equation}
Inserting Eqs.~(\ref{eq:I1_BI}), (\ref{eq:I2_BI_case1}), and (\ref{eq:I2_BI_case2}) into Eq.~(\ref{eq:S3_BI}), we obtain 
\begin{equation}
S^{(A)} = 
{2\over (2-{1\over \mu_A})(1-{2\over \mu_P})} {m_A m_P \over \langle m\rangle} {\mu_P \over \mu_A} N^{-1}
\label{eq:S_BI_case1}
\end{equation}
for $1\leq \mu_A<2$ and $\mu_P>2$ and 
\begin{equation}
S^{(A)} = 
{2 \zeta({2\over \mu_P}) \over 2-{1\over \mu_A}} {m_A m_P \over \langle m\rangle} {\mu_P \over \mu_A} N^{-{2\over \mu_A}}
\label{eq:S_BI_case2}
\end{equation}
for $\mu_A>2$ and $1\leq \mu_P<2$.  If we introduce the scaling exponents $\theta_A$ and $\theta_P$ as 
\begin{equation}
S^{(A)} \sim N^{\theta_{A}} \ {\rm and } \ S^{(P)} \sim N^{\theta_{P}},
\end{equation}
they are given by 
\begin{eqnarray}
(\theta_{A},\theta_{P}) &=& \left(-1, -{2\over \mu_P}\right)   \ {\rm for } \ 1\leq \mu_A<2, \mu_P>2, \nonumber\\ 
(\theta_{A},\theta_{P}) &=& \left(-{2\over \mu_A},-1\right)   \ {\rm for } \  \mu_A>2, 1\leq\mu_P<2.  
\label{eq:theta_BI}
\end{eqnarray}

\begin{figure}
\includegraphics[width=0.9\columnwidth]{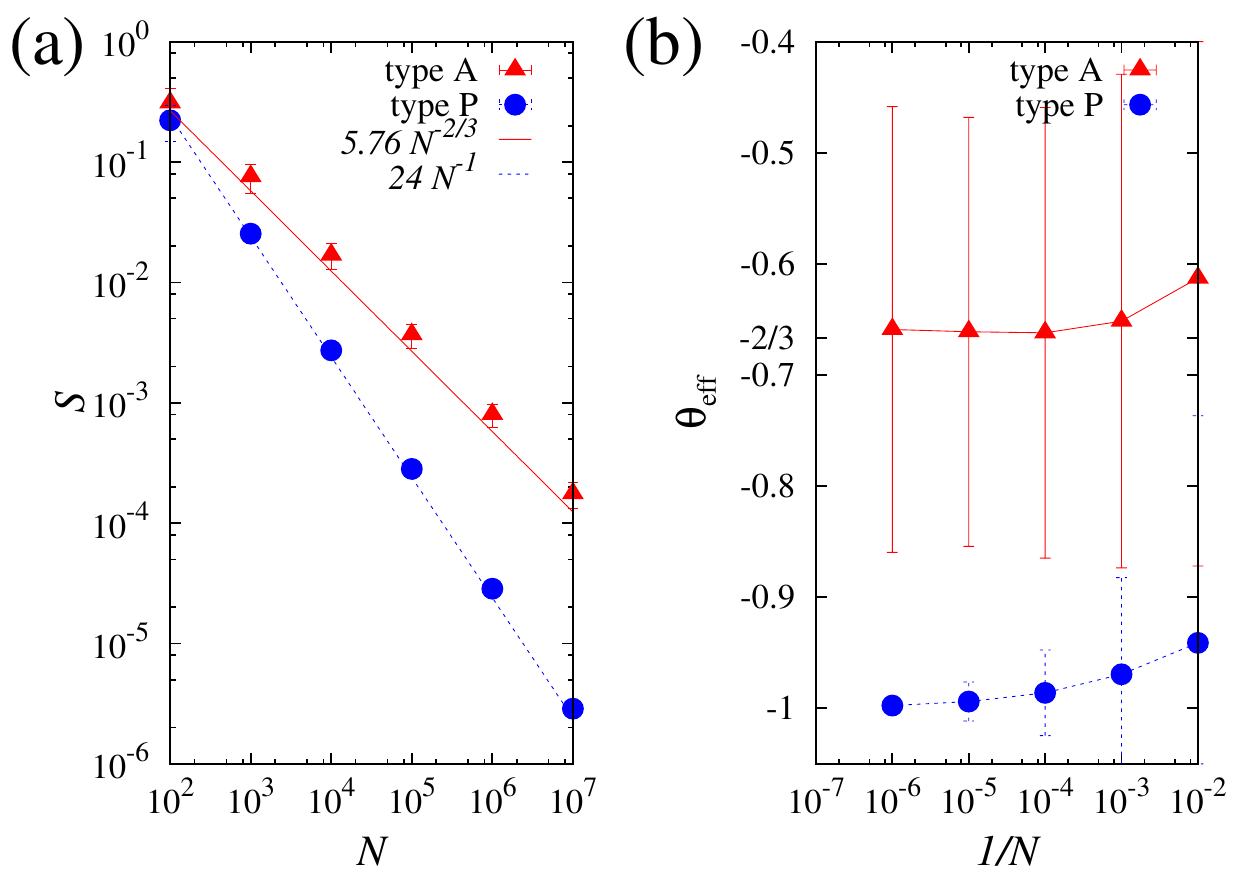}
\caption{Scaling of nestedness in bipartite networks. (a) Logarithmic plot of $S$ versus system size $N$ for type-$A$ and type-$P$ nodes in BA-type bipartite networks of $\rho_A={3\over 4}, \rho_P={1\over 4}, m_A = 2$, and $m_P=3$. Data points are from simulation results, and lines are from analytic results. $S^{(A)}={8\over 5}\zeta({4\over 3}) N^{-{2\over 3}}\simeq 5.76 N^{-{2\over 3}}$ and $S^{(P)}=24N^{-1}$  from Eqs.~(\ref{eq:S_BI_case1}) and (\ref{eq:S_BI_case2}), respectively. (b) Effective exponent $\theta_{\rm eff}$ obtained numerically from simulation results plotted as a function of the inverse of the system size $1/N$. The analytic results $\theta_A=-{2\over \mu_A}=-{2\over 3}$ for type-$A$ nodes and $\theta_P=-1$ for  type-$P$ nodes  are also shown for comparison. The error bars represent the standard deviation.}
\label{fig:bi}
\end{figure}

The most remarkable feature of the nestedness of BA-type bipartite networks is that its scaling behavior is affected by the fraction of each type of node and the number of initial links of new nodes,  $\rho_A, \rho_P, m_A$, and $m_P$.  As a result, the nestedness turns out to behave  differently for the two types of nodes. For instance, when $\mu_A>2$ and $1\leq \mu_P<2$, $\rho_A m_A > \rho_P m_P$, and one can expect that  type-$A$ nodes have more chance to share partners of type $P$ due to the smaller pool of their potential partners  than type-$P$ ones do.  Our results show that the nestedness of type-$A$ nodes is, indeed, much higher than that of type-$P$ nodes: $S^{(A)} \sim N^{-{2\over \mu_A}}\gg S^{(P)} \sim N^{-1}$. The simulation results of $S^{(A)}$ and $S^{(P)}$ for BA-type bipartite networks with $\mu_A=3$ and $\mu_P={3\over 2}$  are shown in Fig.~\ref{fig:bi} and are in good agreement with the analytic predictions of Eq.~(\ref{eq:theta_BI}). 

\section{Summary and Discussion}
\label{sec:discussion}

Motivated by the nestedness observed in plant-pollinator mutualistic networks, we derived in this work the scaling behavior of the nestedness in several random network models. While the nestedness decays with increasing system size, vanishing in the thermodynamic limit, its scaling behavior exhibits a crucial dependence on the connectivity patterns, implying a variation of nestedness with network structure in finite-size systems. The nestedness becomes much larger in strongly heterogeneous networks, those with a degree exponent between 2 and 3, than in more homogeneous networks. Therefore, the impact of nestedness on the robustness of mutualistic networks~\cite{thebault10} can be better understood by relating it to the stability of scale-free networks with hub nodes~\cite{albert02}. In the BA-type bipartite networks, the nestedness  is shown to very sensitively depend on the  global characteristics such as the fraction of different types of nodes and the number of initial links of new nodes, suggesting the possibility that  the species richness and the details of evolutionary patterns in ecological systems are not random, but are remodeled through evolutionary selection for structural and functional robustness. Our results for model networks can be used as a reference for exploring empirical networks in the disciplines of biology, ecology, and economic ecosystems, enabling one to identify novel structural characteristics and to better understand their design and working principles.

\begin{acknowledgments}
We thank Jae Sung Lee for useful discussion. This work was supported by Inha University Research Grant (INHA-40880).
\end{acknowledgments}

\appendix
\section{Computation of nestedness in the static model}
\label{sec:appendix_static}

In the static model, the connection probability  $f_{i\ell}$ in Eq.~(\ref{eq:f}) is factorized as in Eq.~(\ref{eq:factorized_static}) only for $\ell\gg c(i)$ with 
\begin{equation}
c(i) = \left({2NK \over \zeta_N(\alpha)^2}\right)^{1\over \alpha} {1\over i},
\label{eq:ci}
\end{equation}
which is of order $N^{2-{1\over \alpha}}{1\over i}$. In this appendix, we derive the nestedness of static-model networks by considering the fact that $f_{i\ell}\simeq 1$ for $\ell\ll c(i)$ when $\alpha\geq {1\over 2}$ and show that, as far as its leading behavior in the large-$N$ limit is concerned, $f_{i\ell}$ can be approximated as in  Eq.~(\ref{eq:factorized_static}).

The expected degree $\langle k_i\rangle$ has been derived in Refs.~\cite{lee04} and \cite{jslee06} by applying the Euler-Maclaurin formula as
\begin{eqnarray}
\langle k_i \rangle &\simeq& \int_1^N d\ell (1-e^{-2NKP_i P_\ell}\simeq 2K(1-\alpha) \left({N\over i}\right)^\alpha + \mathcal{A}, \nonumber\\
\ {\rm with} &&  \ \mathcal{A} = \left\{
\begin{array}{ll}
\mathcal{O}\left({1\over N}\left({N\over i}\right)^\alpha\right) &\ {\rm for} \ i\gg c(1) \\
\mathcal{O}\left(N^{1-{1\over \alpha}} {N\over i}\right)&\ {\rm for} \ i\ll c(1).
\end{array}
\right.
 \label{eq:expectedk}
\end{eqnarray}
The remainder term $\mathcal{A}$ is negligible compared with the leading term in the limit $N\to\infty$. 
The leading term in Eq.~(\ref{eq:expectedk}) is equal to the one that would be obtained by inserting $f_{i\ell}\approx 2NKP_i P_\ell$ into $\langle k_i\rangle =\sum_\ell f_{i\ell}$ in the whole range of $\ell$ without regard to $i$ or $\alpha$ because the sum is dominantly contributed to by $f_{i\ell}$'s with $\ell\gg c(i)$~\cite{lee04,jslee06}.

To derive the scaling exponent $\theta$ in Eq.~(\ref{eq:Sasym})~\cite{approximation},  we consider the upper and the lower bounds of $f_{i\ell}$,  $(1-{1\over e}) \bar{f}_{i\ell}\leq f_{i\ell}\leq \bar{f}_{i\ell}$, where 
\begin{eqnarray}
\bar{f}_{i\ell} &=& \bar{f}(2NKP_iP_\ell), \ \ {\rm with} \ \nonumber \\
\bar{f}(x) &=&  \left\{
\begin{array}{ll}
x  &  {\rm for} \ x \leq 1\\
1 &  {\rm for} \ x > 1.
\end{array}
\right.
\label{eq:barf}
\end{eqnarray}
Also, we introduce a node-dependent quantity $g_i$, which is the restricted sum of the links of neighboring nodes of $i$, defined as
\begin{equation}
g_i = \sum_{j=1}^{i-1}\sum_{\ell=1}^N f_{i\ell}f_{j\ell}.
\label{eq:gi}
\end{equation}
Then, the nestedness $S$ is represented as
\begin{equation}
S = {2\over N(N-1)} \sum_{i=1}^N {g_i \over \langle k_i\rangle}.
\label{eq:Sgi}
\end{equation}
We note the difference between the quantity ${g_i\over \langle k_i\rangle}={1\over \langle k_i\rangle}\sum_{\ell\in {\rm n.n.}(i)} \sum_{j=1}^{i-1} f_{j\ell}$ and  the expected mean neighboring degree $k_{\rm nn}(i) = {1\over \langle k_i\rangle}\sum_{\ell\in {\rm n.n.}(i)}\sum_{j=1}^N f_{j\ell}$~\cite{satorras01,jslee06}.  In ${g_i\over \langle k_i\rangle}$, only the links incident on nodes $j$ with $j<i$, which are expected to have larger degrees than node $i$, are counted. Therefore, ${g_i\over \langle k_i\rangle}$, in general increases with increasing $i$ while $k_{\rm nn}(i)$ is constant or decreases with increasing $i$ in scale-free networks with small degree exponents~\cite{park03,jslee06}. 

Computing $g_i$'s and $S$ by using $\bar{f}$'s and $(1-{1\over e})\bar{f}$'s yields the same results as far as the scaling exponent $\theta$ in $S\sim N^\theta$ is concerned.  Therefore, for the static model, we evaluate $g_i$ by using
\begin{equation}
g_i = \sum_{j=1}^{i-1}\sum_{\ell=1}^N \bar{f}_{i\ell} \bar{f}_{j\ell}.
\label{eq:gi2}
\end{equation}

\begin{figure}
\includegraphics[width=0.9\columnwidth]{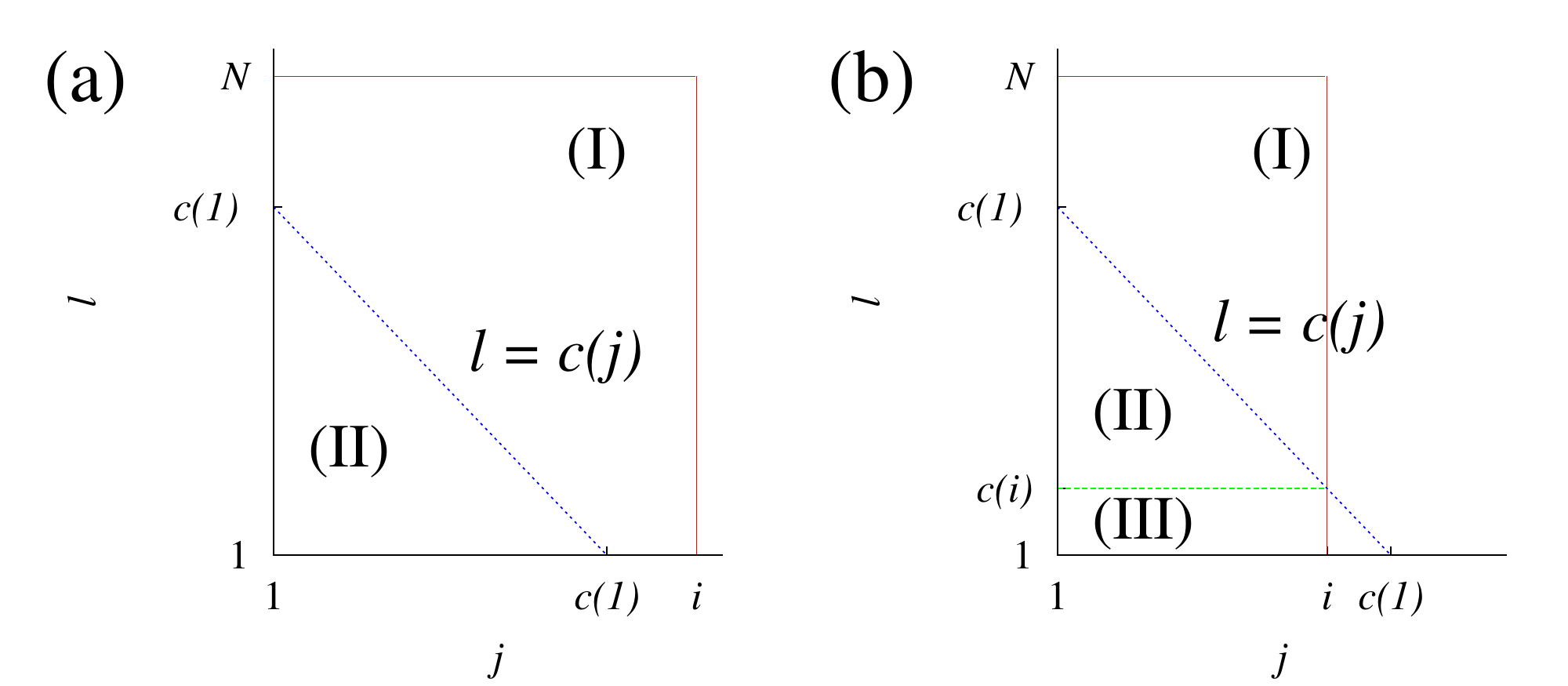}
\caption{Disjoint regions  in  $R = \{(j,\ell)| j=1,2,\ldots, i-1, \ell=1,2,\ldots, N\}$, in which  $\bar{f}_{j\ell}$ and $\bar{f}_{i\ell}$ take different forms. (a) Case of $i\gg c(1)$.  Two regions (I) and (II) should be considered, which are $R_{II} = \{(j,\ell)\in R| 1\leq j <c(1), 1\leq \ell<c(j)\}$ and $R_{I} = R - R_{II}$.  (b) Case of $i\ll c(1)$.  Three regions (I), (II), and (III) should be considered, which are $R_{III} = \{(j,\ell)\in R| 1\leq j <i, 1\leq \ell< c(i)\}$, $R_{II} = \{(j,\ell)\in R| 1\leq j <i, c(i)\leq \ell< c(j)\}$, and $R_{I} = \{(j,\ell)\in R| 1\leq j <i, c(j)\leq \ell\leq N\}$. }
\label{fig:region}
\end{figure}

When $0\leq \alpha<{1\over 2}$, considering $\zeta_N(\alpha)\sim N^{1-\alpha}$, one can see that both  $2NKP_i P_\ell$ and $2NKP_jP_\ell$ are much less than $1$  in the entire region $1\leq j\leq i-1$ and $1\leq \ell\leq N$, leading to 
\begin{eqnarray}
g_i &=& \sum_{\ell=1}^N \sum_{j=1}^{i-1} (2NK)^2 P_i P_j P_\ell^2 \nonumber \\
&=&(2NK)^2 P_i {\zeta_N(2\alpha)\zeta_i (\alpha)\over \zeta_N(\alpha)^3}\nonumber \\
&\simeq& 2NK P_i {2K (1-\alpha)^2 \over 1-2\alpha}\left({i\over N}\right)^{1-\alpha}.
\label{eq:gi_case1}
\end{eqnarray}
When ${1\over 2}<\alpha<1$, we see that $2NKP_iP_\ell>1$ for $\ell<c(i)$ and $2NKP_jP_\ell >1$ for $\ell<c(j)$, with $c(i)$ in Eq.~(\ref{eq:ci}), so we consider separately the cases of $c(i)< 1$ and $c(i)>1$ to compute Eq.~(\ref{eq:gi2}).  
If $c(i)<1$ or equivalently, $i> c(1)= ({2NK\over \zeta_N(\alpha)^2})^{1\over \alpha}$, it holds that $\bar{f}_{i\ell}=2NKP_iP_\ell$ for all the considered values of $\ell$ and $j$.  $\bar{f}_{j\ell} $ behaves differently in regions  (I) and (II) shown in Fig.~\ref{fig:region}(a):  $\bar{f}_{j\ell}=2NKP_jP_\ell$ in region (I) and $\bar{f}_{j\ell}=1$ in region (II). Therefore, $g_i$ is computed as 
\begin{eqnarray}
g_i &=& \sum_{\ell=1}^{N} \sum_{j=1}^{i-1} (2NK)^2 P_iP_j P_\ell^2 \nonumber \\
&&+ \sum_{j=1}^{c(1)}\sum_{\ell=1}^{c(j)} \left\{
2NKP_iP_\ell - (2NK)^2 P_iP_jP_\ell^2\right\} \nonumber \\
&\simeq & \ 2NK P_i [2K (1-\alpha)^2 \zeta(2\alpha)] N^{2\alpha-1} \left({i\over N}\right)^{1-\alpha} \\
&&- 2NKP_i (2K(1-\alpha)^2)^{1/\alpha} N^{1+\alpha-{1\over \alpha}}.
\label{eq:gi_case2}
\end{eqnarray}
Note that among the leading terms of the two sums given in the last line, the first one is dominant.

When $c(i)>1$ or equivalently $i< c(1)$, one should consider the three regions shown in Fig.~\ref{fig:region}(b), for which one finds that  $\bar{f}_{i\ell}=2NKP_jP_\ell$ and $\bar{f}_{j\ell}=2NKP_j P_\ell$ in region (I),  $\bar{f}_{i\ell}=2NKP_iP_\ell$ and $\bar{f}_{j\ell}=1$ in region (II), and $\bar{f}_{i\ell}=\bar{f}_{j\ell}=1$ in region (III).  Then, $g_i$ is given by 
\begin{equation}
\begin{split}
g_i =& \sum_{j=1}^{i-1} \left[
\sum_{\ell=1}^{c(i)} 1 + 
\sum_{\ell=c(i)}^{c(j)} 2NKP_i P_\ell +\right.\\
&+\left.\sum_{\ell=c(j)}^N (2NK)^2 P_i P_j P_\ell^2\right]  \\
\simeq& \ 2NKP_i [2K(1-\alpha)^2]^{{1\over\alpha}-1} \\
\times& {(1-\alpha)(2\alpha+1)\over 2\alpha-1} N^{2-{1\over\alpha}}\left({i\over N}\right)^\alpha, 
\end{split}
\label{eq:gi_case3}
\end{equation}
where the three sums all contribute to the leading behavior of $g_i$ given in the last line. 

Using $g_i$ given in Eqs.~(\ref{eq:gi_case1}) and the expected degree $\langle k_i \rangle$ in Eq.~(\ref{eq:expectedk}), we see that the nestedness $S$ is given as in Eq.~(\ref{eq:S_static_case1})
for $0\leq \alpha<{1\over 2}$.  
When ${1\over 2}<\alpha<1$ or equivalently $2<\gamma<3$, the nestedness is evaluated by using Eqs.~(\ref{eq:expectedk}), (\ref{eq:gi_case2}), and (\ref{eq:gi_case3}) as  
\begin{eqnarray}
S &=& {2\over N(N-1)} \sum_{i=1}^{c(1)} [2K(1-\alpha)^2]^{{1\over\alpha}-1} \nonumber\\
&&\times 
{(1-\alpha)(2\alpha+1)\over 2\alpha-1}
N^{2-{1\over\alpha}}\left({i\over N}\right)^{\alpha}\nonumber \\
&+& {2\over N(N-1)} \sum_{i=i_*}^N 2K(1-\alpha)^2 \zeta(2\alpha)N^{2\alpha-1} \left({i\over N}\right)^{1-\alpha}\nonumber\\
&\simeq & {4K(1-\alpha)^2 \over 2-\alpha} \zeta(2\alpha) N^{2\alpha-2},
\label{eq:S_case2}
\end{eqnarray}
where we used the fact that the sums in the range $i\in [c(1),N]$ make the dominant contribution to $S$.


\begin{references}
\bibitem{LS} Properties of real-world ecological networks are too variable to fix the scaling relation between $N$ and $L$~\cite{pascual05}.
\bibitem{approximation} It has been shown that the approximation of the double sums in Eq.~(\ref{eq:S2})  by integrals may create error terms of the same order as the leading term~\cite{jslee06}.
\end{references}
\end{document}